\documentclass[useAMS,usenatbib]{mn2e}
\usepackage{graphicx}
\renewcommand{\vec}[1]{\mbox{\boldmath $#1$}}

\title[Predicting the dependence of differential rotation on temperature and rotation]
{Differential rotation of main-sequence dwarfs: predicting the
dependence on surface temperature and rotation rate}
\author[L.\,L.\,Kitchatinov and S.\,V.\,Olemskoy]
{L.\,L.\,Kitchatinov$^{1,2}$\thanks{E-mail: kit@iszf.irk.ru}
and S.\,V.\,Olemskoy$^{1}$ \\
$^{1}$Institute for Solar-Terrestrial Physics, PO Box 291, Irkutsk,
664033, Russia\\
$^{2}$Pulkovo Astronomical Observatory, St. Petersburg, 196140,
Russia
 }

\begin{document}

\date{Accepted .... Received ...; in original form ...}

\pagerange{\pageref{firstpage}--\pageref{lastpage}} \pubyear{2011}

\maketitle

\label{firstpage}

\begin{abstract}
Gyrochronology and recent theoretical findings are used to reduce
the number of input parameters of differential rotation models. This
eventually leads, after having fixed our turbulence model
parameters, to a theoretical prediction for the surface differential
rotation as a function of only two stellar parameters - surface
temperature and rotation period - that can be defined
observationally. An analytical approximation for this function is
suggested. The tendency for the differential rotation to increase
with temperature is confirmed. The increase is much steeper for late
F-stars compared to G- and K-dwarfs. Slow and fast rotation regimes
for internal stellar rotation are identified. A star attains its
maximum differential rotation at rotation rates intermediate between
these two regimes. The amplitude of the meridional flow increases
with surface temperature and rotation rate. The structure of the
flow changes considerably between cases of slow and fast rotation.
The flow in rapid rotators is concentrated in the boundary layers
near the top and bottom of the convection zone with very weak
circulation in between.
\end{abstract}
\section{Introduction}
Rotation of the sun and other stars with external convection zones
is not uniform. The rotation period depends on latitude and typically
increases from the equator to the poles \citep{DSB96,Cam03,Cam07}. The
phenomenon of differential rotation is closely related to magnetic
activity and has long been a subject of theoretical studies. The
currently leading theoretical concept pioneered by \citet{L41}
explains differential rotation by the interaction between convection
and rotation. Convective motions in a rotating star are disturbed by
Coriolis force. Back reaction disturbs rotation to make it not
uniform.

Theoretical models reproduce the observed rotation of the sun quite
closely \citep{KR05,MBT06}. In the case of the sun, however, the
differential rotation theory has to explain what is already known
from helioseismology. The detailed helioseismological picture of the
internal solar rotation \citep{WBL97,Sea98} left no opportunity for
theoretical predictions. The theory then has to test its predictive
ability with other stars, and this is very tempting to do in view of
the rapid development of asteroseismology \citep{CH10}. Some
computations of differential rotation as a function of stellar mass
and rotation rate have been already attempted \citep{KR99} and were
even to some extent supported by observations \citep{Bea05}. A
systematic study of stellar differential rotation was, however, not
possible because the stellar parameters on which it depends are too
many. The resulting differential rotation depends on mass, chemical
composition (metallicity), rotation period, and age of a star.
Exploring four-dimensional parametric space is an unbearable task.

The situation has changed recently. First, the development of a new
branch of astronomy named Gyrochronology has led to the
establishment of a functional relation between stellar mass, age,
and rate of rotation \citep{B03,B07,B10,Cea09,MMS09}. Second,
theoretical modeling shows that main-sequence stars of different
mass and metallicity have (almost) the same differential rotation
when their surface temperatures coincide (\citeauthor{KO11}
\citeyear{KO11}, hereafter KO11). In other words, the dependencies
on mass and metallicity can be combined into a common dependence on
effective temperature. These two developments reduce the number of
independent parameters on which differential rotation depends by
two. The dependence on the remaining two parameters can well be
explored numerically.

In our preceding publication (KO11), a new mean-field model for
differential rotation was presented. Performance of the model was
tested by applying it to the sun and individual stars including two
rapid rotators (AB~Dor and LQ~Hya) and two moderate rotators
($\epsilon$~Eri and $\kappa^1$~Ceti) whose rotation has been studied
well observationally \citep{DC97,Kea04,Cea06,Wea07}. The temperature
dependence of differential rotation of young stars observed by
\citet{Bea05} was modeled. This new paper reports and summarizes the
results of extensive computations of stellar differential rotation.
We find that after a solar-like star arrives at the main sequence,
its differential rotation initially increases and then decreases as
the star ages. The rotation period, at which the differential
rotation is maximum, decreases with stellar mass. The dependence on
rotation rate is, however, mild. The differential rotation depends
more strongly on surface temperature, the hotter the star, the
larger it is.  In this paper, we suggest an analytical approximation
for the numerically defined dependence of differential rotation on
rotation period and surface temperature. The theoretical prediction
can be tested by observations if only rotation period, surface
temperature and the surface differential rotation are simultaneously
measured. Dependence of the meridional flow amplitude and structure
on rotation rate and temperature is also studied. The next
Section~\ref{m&m} describes our model and method. Section~\ref{r&d}
presents and discusses the results, which are summarized in
Section~\ref{S}.
\section{Model and method}\label{m&m}
\subsection{The model}
Our model is based on the mean-field theory of differential
rotation. This means that the velocity field ($\vec v$) is split in
its mean ($\vec V$) and fluctuating ($\vec u$) parts,
\begin{equation}
    {\vec v} = {\vec V} + {\vec u},\ \ \ {\vec V} = \langle{\vec
    v}\rangle,\ \ \ \langle{\vec u}\rangle = 0,
    \label{1}
\end{equation}
and equations for the mean flow, $\vec V$, are formulated. The
equations are then solved numerically. The angular brackets in
(\ref{1}) signify an averaging procedure that smoothes out random
fluctuations of turbulent fields. The steady equation for the mean
flow,
\begin{equation}
    \left({\vec V}\cdot{\vec\nabla}\right){\vec V} +
    \frac{1}{\rho}{\vec\nabla}P - {\vec g} =
    \frac{1}{\rho}{\vec\nabla}\cdot{\vec R} ,
    \label{2}
\end{equation}
involves the effects of turbulence via the Reynolds stress tensor,
\begin{equation}
    R_{ij} = -\rho\langle u_i u_j\rangle.
    \label{3}
\end{equation}
In equation (\ref{2}), $\rho$, $P$, and $\vec g$ are the mean
density, pressure, and gravity respectively. Reynolds stress
(\ref{3}) includes two parts representing the turbulent viscosity
and non-viscous flux of momentum,
\begin{equation}
    R_{ij} = -\rho Q^\Lambda_{ij}
    + \rho{\cal N}_{ijkl}\frac{\partial V_k}{\partial r_l} ,
    \label{RS}
\end{equation}
where the summation convention over the repeated subscripts is
assumed, ${\cal N}_{ijkl}$ is the eddy viscosity tensor, and
$Q^\Lambda_{ij}$ represents the non-viscous part of the stress (cf.
Appendix of KO11 for detailed expressions for both parts of the
Reynolds stress (\ref{RS})). The presence of non-viscous flux of
angular momentum in rotating turbulent fluids was called the
\lq$\Lambda$-effect' \citep{R89}. This effect is the principal
driver of differential rotation in our model.

The direction of the angular momentum transport by the
$\Lambda$-effect is controlled by two dimensionless parameters,
$V^{(0)}$ and $H^{(1)}$, of the $Q^\Lambda$-tensor
\begin{eqnarray}
    Q_{ij}^\Lambda &=& \nu_{_\mathrm{T}}\Omega_k\hat{r}_l
    \left( V^{(0)}(\Omega^*)\left( \hat{r}_i\varepsilon_{jkl}
    + \hat{r}_j\varepsilon_{ikl}\right)\right.
    \nonumber \\
    &+& \left. H^{(1)}(\Omega^*) \frac{\left( {\hat{\vec r}}\cdot
    {\vec\Omega}\right) }{\Omega^2}\left(\Omega_i\varepsilon_{jkl}
    + \Omega_j\varepsilon_{ikl}\right) \right)
    \label{Q} ,
\end{eqnarray}
where $\vec\Omega$ is the angular velocity vector, $\hat{\vec r}$ is
the radial unit vector,
\begin{equation}
    \nu_{_\mathrm{T}} = -\frac{\tau\ell^2 g}{15 c_\mathrm{p}}
    \frac{\partial S}{\partial r}
    \label{nu}
\end{equation}
is the eddy viscosity parameter, $S$ is the specific entropy, $g$ is
gravity, $c_\mathrm{p}$ is the specific heat at constant pressure,
$\ell = \alpha_{_\mathrm{MLT}}H_\mathrm{p}$ is the mixing length,
$H_\mathrm{p}$ is the pressure scale height,
\begin{equation}
    \Omega^* = 2\tau\Omega
    \label{CN}
\end{equation}
is the Coriolis number,
\begin{equation}
    \tau = \left( \frac{4c_\mathrm{p}\rho\ell^2 T}{3g\delta
    F}\right)^{1/3}
    \label{tau}
\end{equation}
is the convective turnover time, $\delta F = L(4\pi r^2)^{-1} -
F^\mathrm{rad}$ is the \lq residual' heat flux that convection has
to transport, and $F^\mathrm{rad}$ is the radiative heat flux.

Coriolis number (\ref{CN}) measures the intensity of interaction
between convection and rotation. Its value defines whether the
turbulent eddies are living long enough for rotation to influence
them considerably. The Coriolis number, $\Omega^* =
4\pi\mathrm{Ro}^{-1}$, is reciprocal to the Rossby number
$\mathrm{Ro} = P_\mathrm{rot}\tau^{-1}$; $P_\mathrm{rot}$ is the
rotation period. Explicit expressions for the functions
$V^{(0)}(\Omega^*)$ and $H^{(1)}(\Omega^*)$ of (\ref{Q}) were
derived by \citeauthor{KR95} (1995,2005). In the slow rotation
limit, $\Omega^* \ll 1$, $H^{(1)}(\Omega^*)$ is small compared to
$V^{(0)}(\Omega^*)$ and $V^{(0)}(\Omega^*)$ is negative. This means
that the non-diffusive flux of angular momentum is close to the
inward radial direction. In the opposite limit of rapid rotation,
$\Omega^* \gg 1$, $V^{(0)}(\Omega^*)$ is small compared to
$H^{(1)}(\Omega^*)$. Angular momentum flux is parallel to the
rotation axis and points to the equatorial plane in this case. As
the Coriolis number increases from small to large values, the
angular momentum transport by the $\Lambda$-effect changes smoothly
from radial inward to equatorward and parallel to the rotation axis.
This picture generally agrees with 3D numerical simulations of
\citet{KB08} and \citet{Kea11}.

The differential rotation model of this paper is identical to that
of KO11. A detailed formulation of the model can be found in that
paper and in the references given there. We just outline the model
here.

The azimuthal component of Eq.~(\ref{2}) in spherical coordinates
($r,\theta,\phi$) gives the mean-field equation for the angular
momentum balance,
\begin{equation}
    \mathrm{div}\left( \rho r \sin\theta\langle u_\phi\vec{u}\rangle
    + \rho r^2\sin^2\theta\ \Omega\vec{V}^\mathrm{m}\right) = 0 ,
    \label{11}
\end{equation}
where ${\vec V}^\mathrm{m}$ is the meridional flow velocity. Upon
substitution of an explicit expression for Reynolds stress in
(\ref{11}), an equation for angular velocity can be obtained,
which involves the meridional flow and is thus not closed.

The origin of meridional flow can be seen from the equation, which
can be obtained as the azimuthal component of the curled equation
(\ref{2}),
\begin{equation}
    {\cal D}(\vec{V}^\mathrm{m})\ =\ \sin\theta\ r{\partial\Omega^2\over\partial z}\
    -\ {g\over c_{\rm p} r}{\partial S\over\partial\theta},
    \label{12}
\end{equation}
where $\partial /\partial z = \cos\theta\partial /\partial r -
r^{-1}\sin\theta\partial /\partial\theta$ is the spatial derivative
along the rotation axis. The left part of this equation describes
the viscous drag on the meridional flow
\begin{equation}
    {\cal D}(\vec{V}^\mathrm{m})\ \equiv
    -\varepsilon_{\phi jk} \frac{\partial}{\partial r_j}
    \left(\frac{1}{\rho}\frac{\partial}{\partial r_l}
    \left(\rho\ {\cal N}_{klmn}\frac{\partial V_m}{\partial r_n}
    \right)\right)
    \label{DV}
\end{equation}
(see Appendix in KO11 for the expression for ${\cal
D}(\vec{V}^\mathrm{m})$ in spherical coordinates).

The right part of the meridional flow equation (\ref{12}) includes
two sources of meridional flow: the centrifugal driving of the flow
by non-conservative part of centrifugal force (known also as the \lq
gyroscopic pumping') and the baroclinic driver due to the meridional
gradient of entropy (thermal wind). Each term on the right side of
(\ref{12}) is large compared to the left side. The two terms on the
right, therefore, almost balance each other, this state is known as
the Taylor-Proudman balance. The meridional flow results from a
slight deviation from the balance. The meridional flow attains its
largest velocities near the boundaries of the convection zone
because the stress-free boundary conditions are not compatible with
the Taylor-Proudman balance. Therefore, the Taylor-Proudman balance
has to be violated near the stress-free boundaries where the source
of meridional flow in the right side of (\ref{12}) attains its
largest values \citep{KR99,MH11}.

Direction of the convective heat flux,
\begin{equation}
    F^\mathrm{conv}_i = -\rho T \chi_{ij}\frac{\partial S}{\partial
    r_j} ,
    \label{13}
\end{equation}
depends on the structure of the eddy conductivity tensor,
\begin{equation}
    \chi_{ij} = \chi\delta_{ij} +
    \chi_\|\frac{\Omega_i\Omega_j}{\Omega^2}.
    \label{14}
\end{equation}
The rotationally induced anisotropy of the conductivity (\ref{14})
results in a deviation of the heat flux (\ref{13}) from a radial
direction to the poles, which in turn produces the dependence of
entropy on latitude. This \lq differential temperature' is,
therefore, produced not only by the dependence of thermal
conductivity on latitude \citep{W65} but mainly by the anisotropy of
the conductivity \citep{Rea05}.

The model solves the system of three joint equations governing
distributions of angular velocity, meridional flow, and entropy
inside the convection zone. These steady equations are solved
numerically by the relaxation method\footnote{The numerical solver
can be obtained freely on request.}.

It should be noted that the model relies on a uniform theoretical
basis: all the model needs, including the $\Lambda$-effect, eddy
viscosities and thermal conductivities was derived within the same
quasi-linear approximation of turbulence theory. In particular, the
eddy conductivities (\ref{14}) were derived as fully nonlinear
functions of the Coriolis number:
\begin{equation}
    \chi = \chi_{_\mathrm{T}}\phi (\Omega^*),\ \ \
    \chi_\| = c_\chi\chi_{_\mathrm{T}}\phi_\| (\Omega^*)
    \label{chi}
\end{equation}
\citep{KPR94}. In this equation, $\chi_{_\mathrm{T}}$ is the
isotropic eddy conductivity which would take place under actual
sources of turbulence (actual superadiabaticity) but in non-rotating
fluid. The mixing-length relation
\begin{equation}
    \chi_{_\mathrm{T}} = -\frac{\tau\ell^2 g}{12 c_\mathrm{p}}
    \frac{\partial S}{\partial r},
    \label{chi1}
\end{equation}
defines the conductivity in terms of the entropy gradient. This
gradient is controlled by the solution of the model equations.
Similarly, the eddy viscosity tensor,
\begin{eqnarray}
    {\cal N}_{ijkl} &=& \nu_1\left(\delta_{ik}\delta_{jl}\ +\
    \delta_{jk}\delta_{il}\right)
    \nonumber \\
    &+& \nu_2\big( \delta_{il}\frac{\Omega_j\Omega_k}{\Omega^2} +
    \delta_{jl}\frac{\Omega_i\hat\Omega_k}{\Omega^2} \nonumber \\
    &+&\delta_{ik}\frac{\Omega_j\hat\Omega_l}{\Omega^2} +
    \delta_{jk}\frac{\Omega_i\hat\Omega_l}{\Omega^2} +
    \delta_{kl}\frac{\Omega_i\hat\Omega_j}{\Omega^2} \big)
    \nonumber \\
    &+& \nu_3\delta_{ij}\delta_{kl}
    - \nu_4 \delta_{ij}\frac{\Omega_k\hat\Omega_l}{\Omega^2} +
    \nu_5\frac{\Omega_i\hat\Omega_j\hat\Omega_k\hat\Omega_l}{\Omega^4}
    ,
    \label{N}
\end{eqnarray}
involves the rotationally induced anisotropy and quenching via
dependence of the viscosity coefficients
\begin{equation}
    \nu_n = \nu_{_\mathrm{T}}\phi_n(\Omega^*)\ \ \ \ n=1,2,...5
    \label{nu1}
\end{equation}
on the Coriolis number (\ref{CN}). Explicit expressions for the
viscosity quenching functions, $\phi_n(\Omega^*)$, are given in
\citet{KPR94} and the background viscosity $\nu_{_\mathrm{T}}$ is
defined in terms of the entropy gradient in (\ref{nu}).

In other words, all the eddy transport coefficients are not
prescribed but computed. Such an approach helps to avoid
arbitrariness in prescribing uncertain parameters. Still, two
parameters remained for some time uncertain. First, the
$\Lambda$-effect depends - apart from the well known (nearly
adiabatic) density stratification - also on turbulence anisotropy,
which is uncertain. The contribution of anisotropy was put to zero
in the former version of our model \citep{KR99} but then fixed by
comparison with 3D f-plane simulations of thermal convection
\citep{KR05} and never varied since then. Second, the $c_\chi$
parameter of (\ref{chi}) is uncertain ($c_\chi = 1$ in Kitchatinov
\& R\"udiger 1999). It was found in KO11 that $c_\chi = 1.5$ gives
close agreement with helioseismology. Since then, $c_\chi = 1.5$ is
used for stellar applications also \citep{KUea11} and not varied any
more. It is $c_\chi = 1.5$ in this paper also.

After the two uncertain parameters were fixed, no further tuning is
possible in the mean-field model for differential rotation. In this
sense, the model does not have free parameters.
\subsection{Reducing the number of input parameters}
Input parameters of the model include the rotation period
($P_\mathrm{rot}$) and a set of parameters, all of which can be
taken from a structure model of a (non-rotating) star. We use the
{\sl EZ} code of stellar evolution by \citet{P04} to specify the
structure of a main-sequence star of given mass, age, and
metallicity. The four basic input parameters - rotation period,
mass, age, and metallicity - are, however, not independent.

\begin{figure}
  \resizebox{\hsize}{!}{\includegraphics{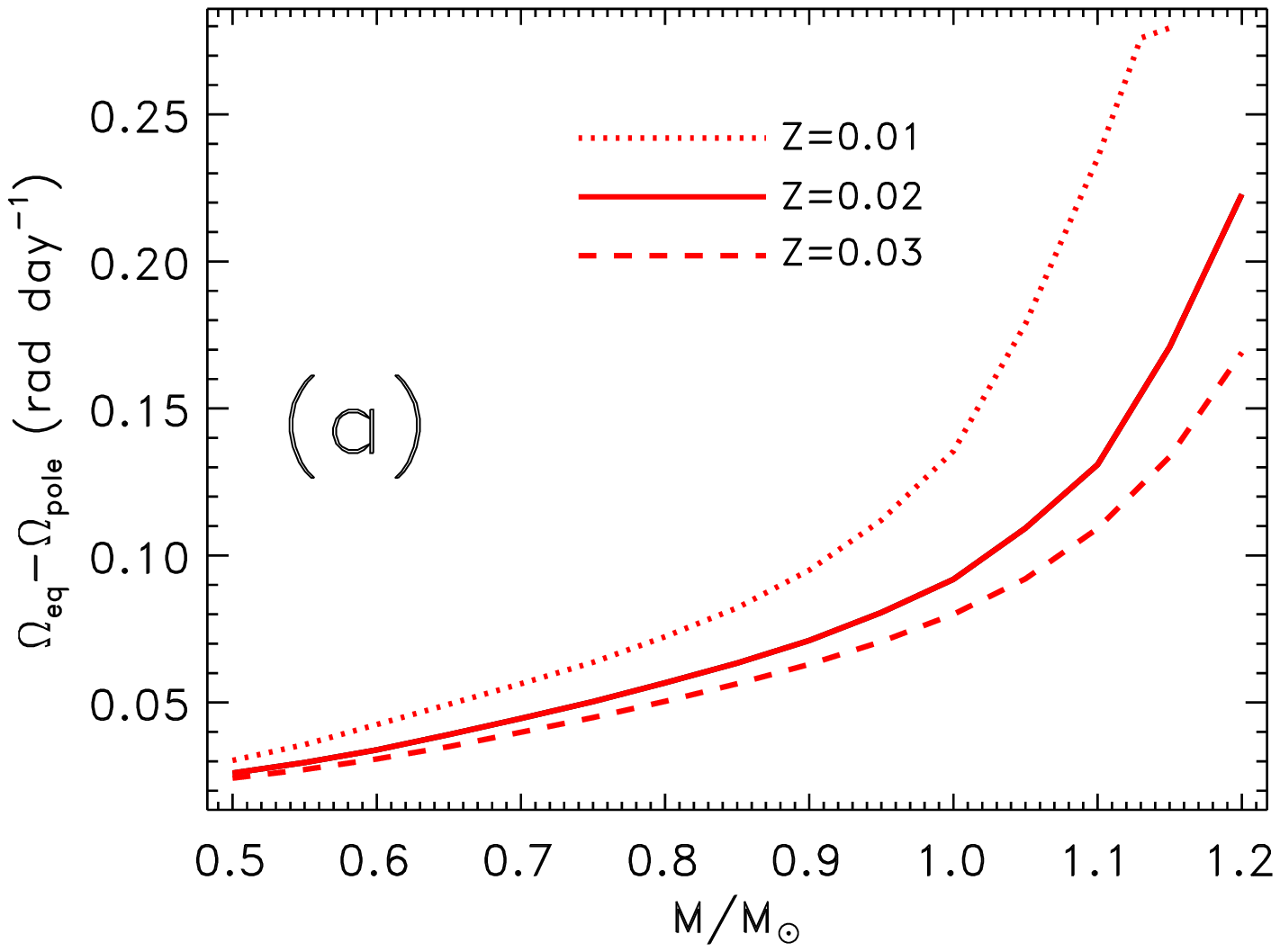}}\\[0.2truecm]
  \resizebox{\hsize}{!}{\includegraphics{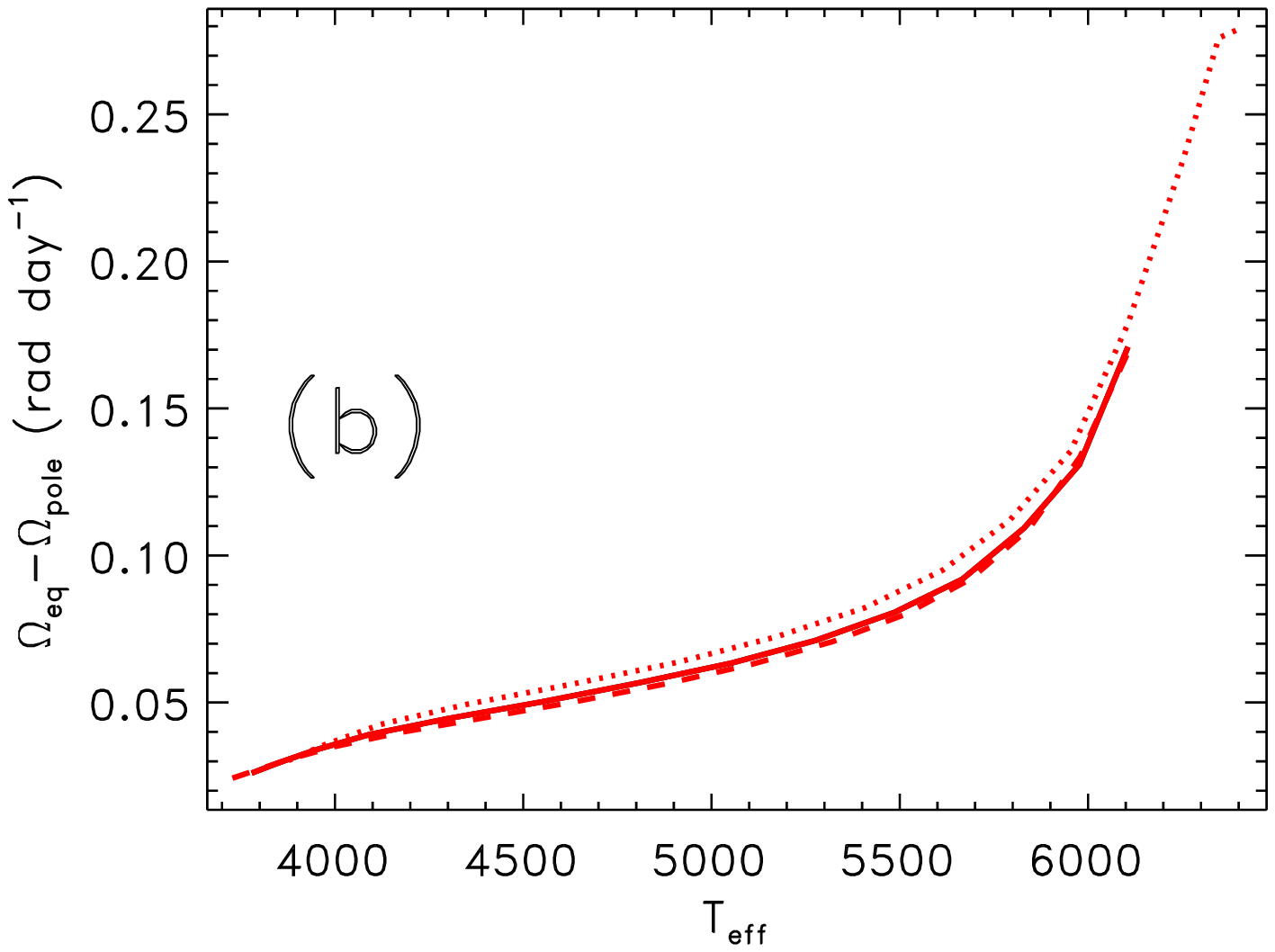}}
  \caption{Surface differential rotation as function of stellar mass
    (a) and surface temperature (b) after our model. Lines of
    different styles correspond to different metallicities $Z$.
    Computations were performed for 1\,Gyr old stars with
    $P_\mathrm{rot} = 10$\,days.
        }
  \label{Z}
\end{figure}

Figure~\ref{Z} shows the surface differential rotation computed with
our model for stars of different mass and metallicity. All stars are
at the age of 1\,Gyr. A rotation period $P_\mathrm{rot} = 10$ days
was prescribed. The dependence on metallicity for stars of given
mass is quite pronounced. However, the dependence almost disappears
when differential rotation is considered as a function of surface
temperature. The same scaling was found in KO11 for ZAMS stars with
$P_\mathrm{rot} = 1$\,day. We conclude that the values of surface
differential rotation computed for stars of different masses and
metallicities but fixed rotation period lie on a single line when
plotted as a function of the effective temperature
($T_\mathrm{eff}$). A number of other computations were done to
confirm this temperature-scaling for stars with not too low
metallicities (Population I stars). The scaling reduces the number
of input parameters by one: the mass and metallicity dependencies
are combined into a common dependence on temperature.

Further reduction is achieved by using Gyrochronology. The rotation
rate of solar- and late-type stars is a function of $T_\mathrm{eff}$
and age ($t$). Gyrochronology generalizes the \citet{S72} law,
$P_\mathrm{rot} \propto t^{0.5}$, by specifying the proportionality
constant in this law as a function of $T_\mathrm{eff}$ or other
equivalent parameter. We use the empirical relation of \citet{B07},
\begin{equation}
    P_\mathrm{rot} = a t^n\left(B-V -c\right)^b \ \mathrm{days},
    \label{4}
\end{equation}
with $n = 0.519$, $a = 0.773$, $b = 0.601$, $c = 0.4$, to specify
the rotation period of a star of given $B - V$ colour and age $t$
measured in Myr. The relation does not apply to very young rapidly
rotating stars of the so-called \lq C-sequence' \citep{B07}. The
structure of a star, however, changes little over its C-sequence
life. Therefore, the relation (\ref{4}) can still be used for our
purposes even for very young stars. We used the tables of
colour-temperature relations and the interpolation code of
\citet{VC03} to compute $B-V$ colour of Eq.~(\ref{4}).

In this paper we fix the metallicity $Z=0.02$ and compute
differential rotation for main-sequence dwarfs of different masses
and ages. The computations cover the mass range from $0.5M_\odot$ to
$1.3M_\odot$. The stellar masses were spanned by $0.05M_\odot$ in
the range of $0.5 \leq M/M_\odot \leq 1.1$. The mass dependence of
differential rotation steepens for more massive stars. Accordingly,
the span was reduced to $0.025M_\odot$ for the range of $1.1 \leq
M/M_\odot \leq 1.3$. For a given mass, structure models were
generated for a sequence of ages. Then, rotation periods of
Eq.~(\ref{4}) were estimated for every age and corresponding
structure, and a sequence of differential rotation models was
computed. This gives the differential rotation as a table-given
function of rotation period and temperature. The surface
differential rotation is then approximated by a neural network with
one hidden layer \citep{C98}:
\begin{equation}
    \Delta\Omega = W_0 + \sum\limits_{n=1}^{N}
    W_n\tanh\left( w_n + w^T_nT_\mathrm{eff} +
    w^P_n P_\mathrm{rot}\right),
    \label{5}
\end{equation}
where $\Delta\Omega = \Omega_\mathrm{eq} - \Omega_\mathrm{pole}$ is
the angular velocity difference between the equator and pole. The
convenience of this approximation is related to i) its universality:
any continuous function can be represented by one hidden layer
network to an arbitrary accuracy given sufficiently large $N$, and
ii) availability of reliable numerical routines for tuning the
parameters of the network (\ref{5}). Using the regression formula
(\ref{5}), a theoretical prediction for the surface differential
rotation of a star can be inferred from an observed rotation period
and effective temperature.
\section{Results and discussion}\label{r&d}
\begin{figure}
  \resizebox{\hsize}{!}{\includegraphics{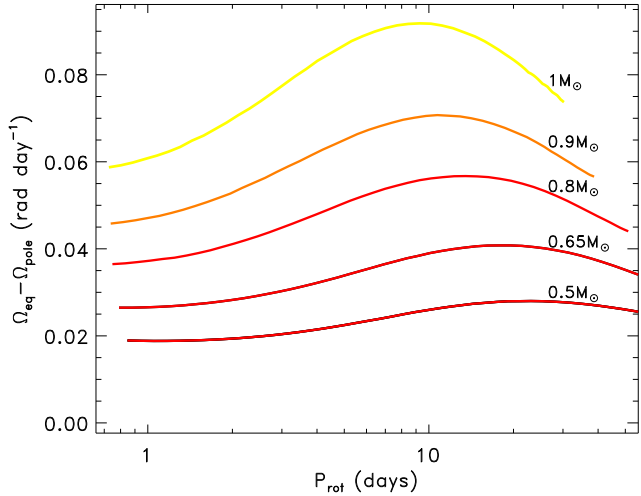}}
  \caption{Surface differential rotation as a function of the rotation period
    for stars of subsolar masses.}
  \label{f1}
\end{figure}

\begin{figure}
  \resizebox{\hsize}{!}{\includegraphics{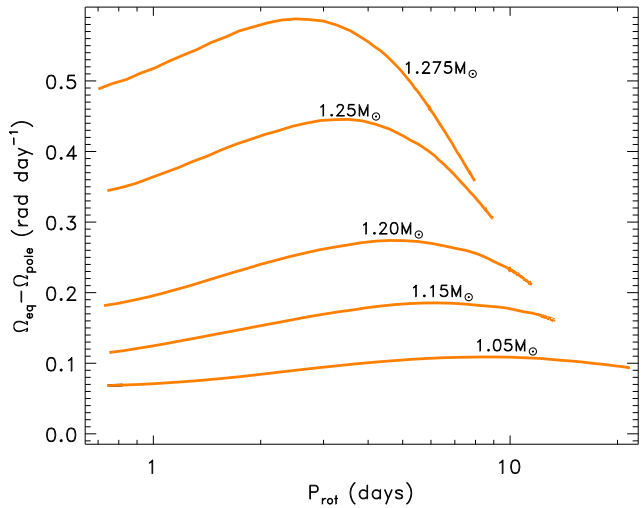}}
  \caption{Surface differential rotation as a function of the rotation period
    for stars more massive than the sun.}
  \label{f2}
\end{figure}

Figures \ref{f1} and \ref{f2} show the computed differential
rotation as a function of the rotation period for a variety of stellar
masses. Lines of these plots were computed for a  certain initial mass
but not for a given stellar structure. The structure changes as a
star ages and rotation period increases.

Dwarf stars do not spin-down beyond a certain maximum rotation
period depending on spectral type (see Fig.\,1 of
\citeauthor{R84}\,\citeyear{R84}). The dependence can be
approximated by the linear function of the $B-V$ colour,
\begin{equation}
    P_\mathrm{rot}^\mathrm{max} = 109 \left( B-V\right) - 43\ \
    \mathrm{day} .
    \label{6}
\end{equation}
Lines of Fig.~\ref{f1} and \ref{f2} are terminated shortly beyond
the maximum rotation period of Eq.\,(\ref{6}).

The dependence of differential rotation on rotation rate is rather
mild. Figures \ref{f1} and \ref{f2} show much stronger dependence on
stellar mass, with more massive and hotter stars having larger
differential rotation. The temperature dependence steepens as
temperature increases. The dependence is in at least qualitative
agreement with differential rotation measurements by Doppler imaging
of \citet{Bea05} (for quantitative comparison of observed and
computed rotation of young stars see Fig.~4 of \citeauthor{K11}
2011). The line of Fig.~\ref{f2} for the hottest star shows the
$\Delta\Omega$ values slightly exceeding the largest differential
rotation observed to date \citep{JD08}.

We used statistics of the results of Fig. \ref{f1} and \ref{f2} and
also the results of computations for stellar masses intermediate
between the lines of these Figures to define the coefficients of the
regression (\ref{5}). Only the results for rotation periods from
slightly less than 1 day to $P_\mathrm{rot}^\mathrm{max}$ of
Eq.~(\ref{6}) were included. Total statistics consists of about 1600
entries from differential rotation computations. The neural network
(\ref{5}) with $N = 6$ gives quite a satisfactory approximation with
a typical error of about 3\%. Coefficients of the regression are
given in Table~1.

\begin{table}
    \label{tab1}
    \caption{Coefficients of the neural network approximation (\ref{5}) for the
    surface differential rotation in terms of the effective temperature and
    rotation period.}
    \begin{tabular}{@{}lcccc}
    \hline
    n & $W_n\left(\frac{\mathrm{rad}}{\mathrm{day}}\right)$ & $w_n$
    & $w^T_n\left(K^{-1}\right)$ &
    $w^P_n\left(\mathrm{day}^{-1}\right)$ \\
    \hline
    $0$ & $0.4521$ & - & - & - \\
    $1$ & $0.1819$ & $-3.573$ & $5.075\times 10^{-4}$ & $-3.118\times 10^{-2}$ \\
    $2$ & $0.1063$ & $2.532$ & $-3.968\times 10^{-4}$ & $0.1139$ \\
    $3$ & $0.6062$ & $-32.78$ & $5.139\times 10^{-3}$ & $-0.2310$ \\
    $4$ & $0.1858$ & $-18.55$ & $2.993\times 10^{-3}$ & $-5.674\times 10^{-2}$ \\
    $5$ & $0.4596$ & $17.90$ & $-2.831\times 10^{-3}$ & $0.4013$ \\
    $6$ & $-0.1293$ & $0.6043$ & $-1.250\times 10^{-4}$ & $1.206\times 10^{-3}$ \\
    \hline
    \end{tabular}
\end{table}

Given the observed rotation period and temperature of a star, a
theoretical prediction for the surface differential rotation can
easily be estimated by using coefficients of this Table. It should
be noted that the estimation applies to single main-sequence dwarfs
only. It cannot be used for giants, close binaries, or pre-main
sequence stars.

For a given mass, the amount of the surface differential rotation of
Fig. \ref{f1} and \ref{f2} varies by about 30\% as the rotation
period changes. This mild variation is, however, worth discussing.
The surface differential rotation increases initially up to a
maximum but then decreases with $P_\mathrm{rot}$. Maxima on all
lines are positioned at the values of the Coriolis number (\ref{CN})
between 10 and 20 (the Coriolis number is estimated for the middle
of convection zone). Stars of smaller mass host slower convection
with larger $\tau$ and attain their largest differential rotation at
smaller rotation rates. The increase of differential rotation with
angular velocity for not too rapid rotation was also found in 3D
simulations of \citet{Bea08}.  The 3D simulations predict, however,
much larger differential rotation compared to our mean-field model.

\begin{figure}
  \resizebox{\hsize}{!}{\includegraphics{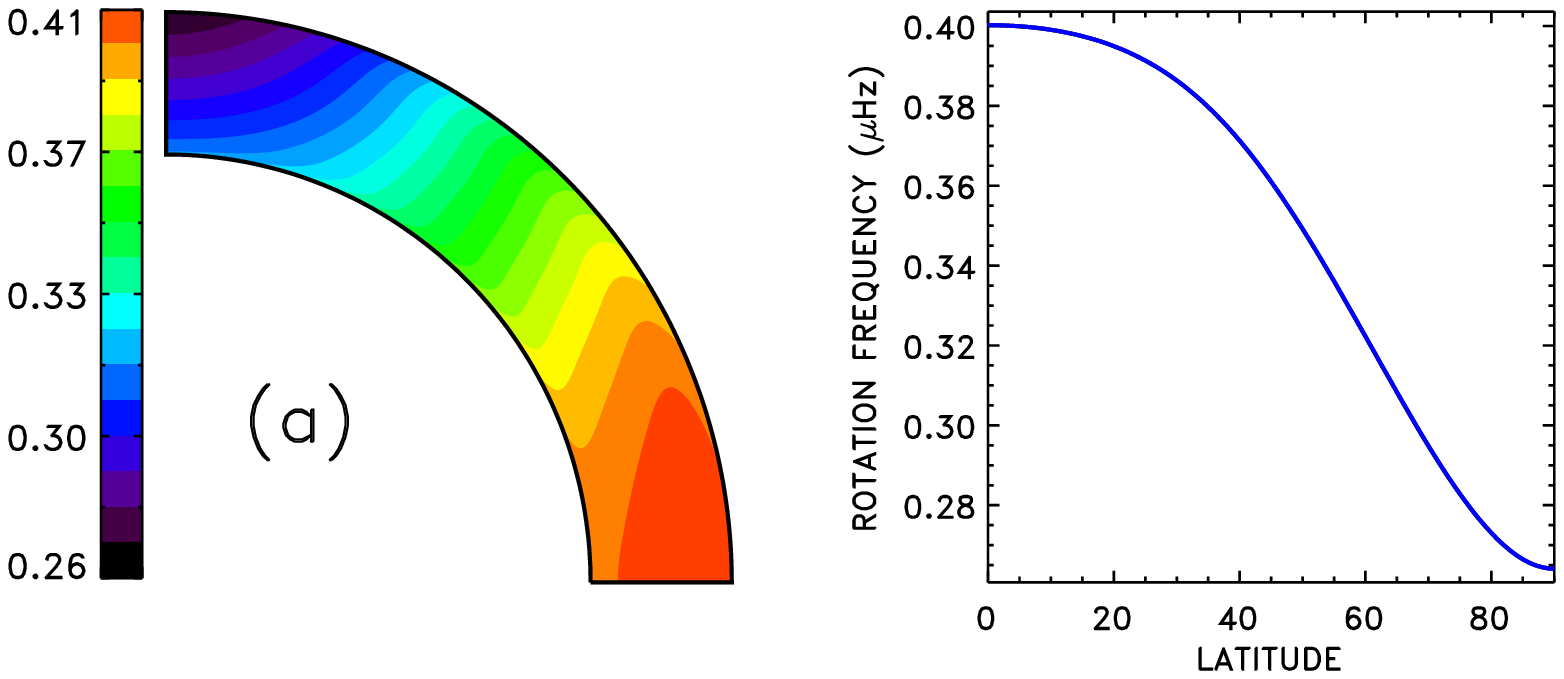}}\\[0.2truecm]
  \resizebox{\hsize}{!}{\includegraphics{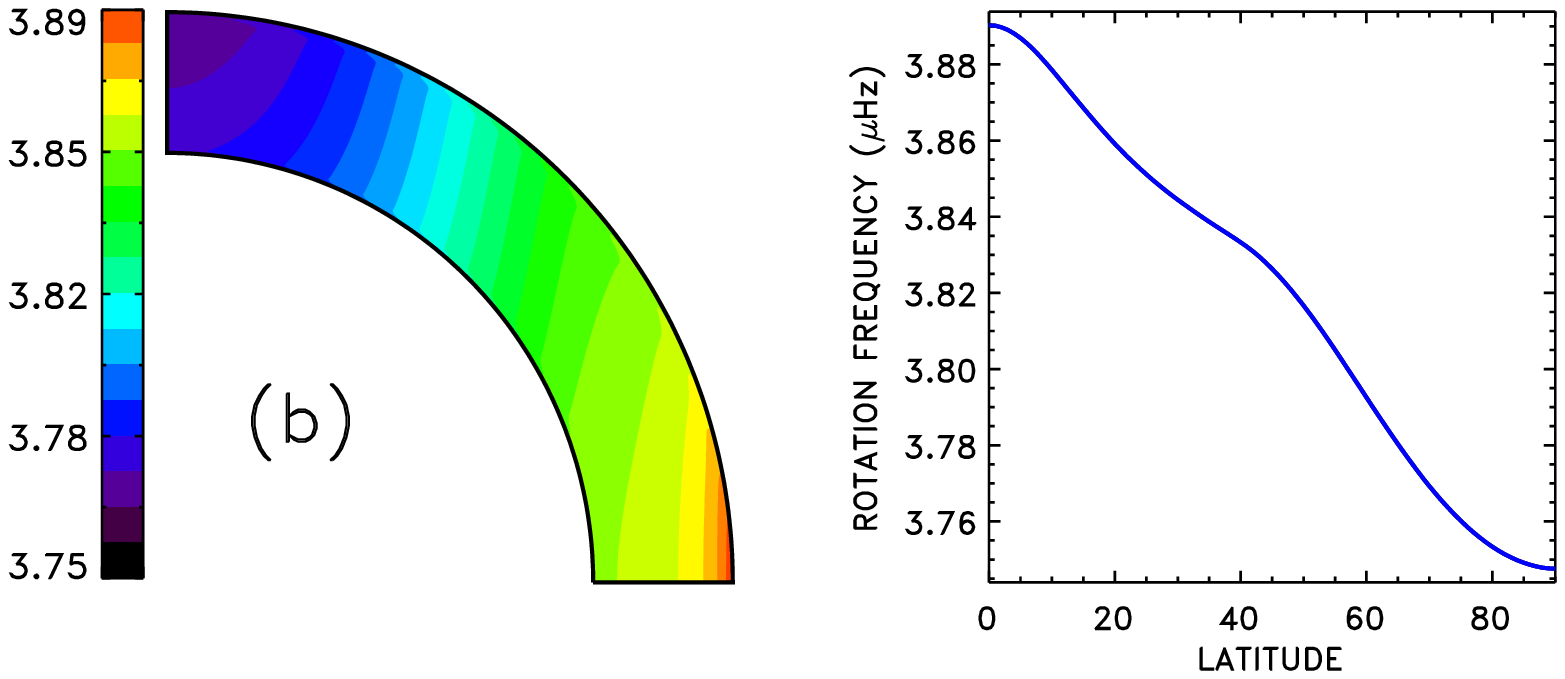}}
  \caption{Angular velocity isolines in the convection zone and
  the angular velocity profile on the surface of $1M_\odot$ stars
    rotating with periods of 30 days (a) and 3 days (b).}
  \label{f3}
\end{figure}

The character of differential rotation changes between cases of slow
($\Omega^* < 10$) and rapid ($\Omega^* > 20$) rotation.
Figure~\ref{f3} shows the computed internal rotation of $1M_\odot$
stars rotating with periods of 30 and 3 days. The slow rotator has a
smooth distribution of angular velocity on the surface and inside
its convection zone. Isorotational surfaces are far from cylinders.
The rapid rotator has thin boundary layers near the top and bottom
of its convection zone. The layers result from violation of the
Taylor-Proudman balance near the stress-free boundaries. The
boundary layers were found in mean-field computations of \citet{D89}
and \citet{KR99} as well as in 3D simulations of \citet{BT02} and
\citet{Bea08}. The profile of angular velocity on the surface of the
rapid rotator has a \lq peculiarity' around the latitude where the
angular velocity isoline tangential to the inner boundary at the
equator arrives at the surface. Isorotational surfaces for the rapid
rotator are much closer to cylinders.

Isorotational surfaces are cylinder-shaped near the equator and
disk-shaped near the axis of rotation in all our computations.
Isorotational surfaces are normal to the equatorial plane due to the
symmetry of angular velocity distribution about the equator. As the
distribution should be regular at the poles, isorotational surfaces
are also normal to the rotation axis. Therefore, it is a general,
though elementary rule, that isorotational surfaces are
cylinder-shaped near the equator and disk-shaped near the poles (for
an alternative opinion see \citeauthor{B09} \citeyear{B09}).

\begin{figure}
  \resizebox{\hsize}{!}{\includegraphics{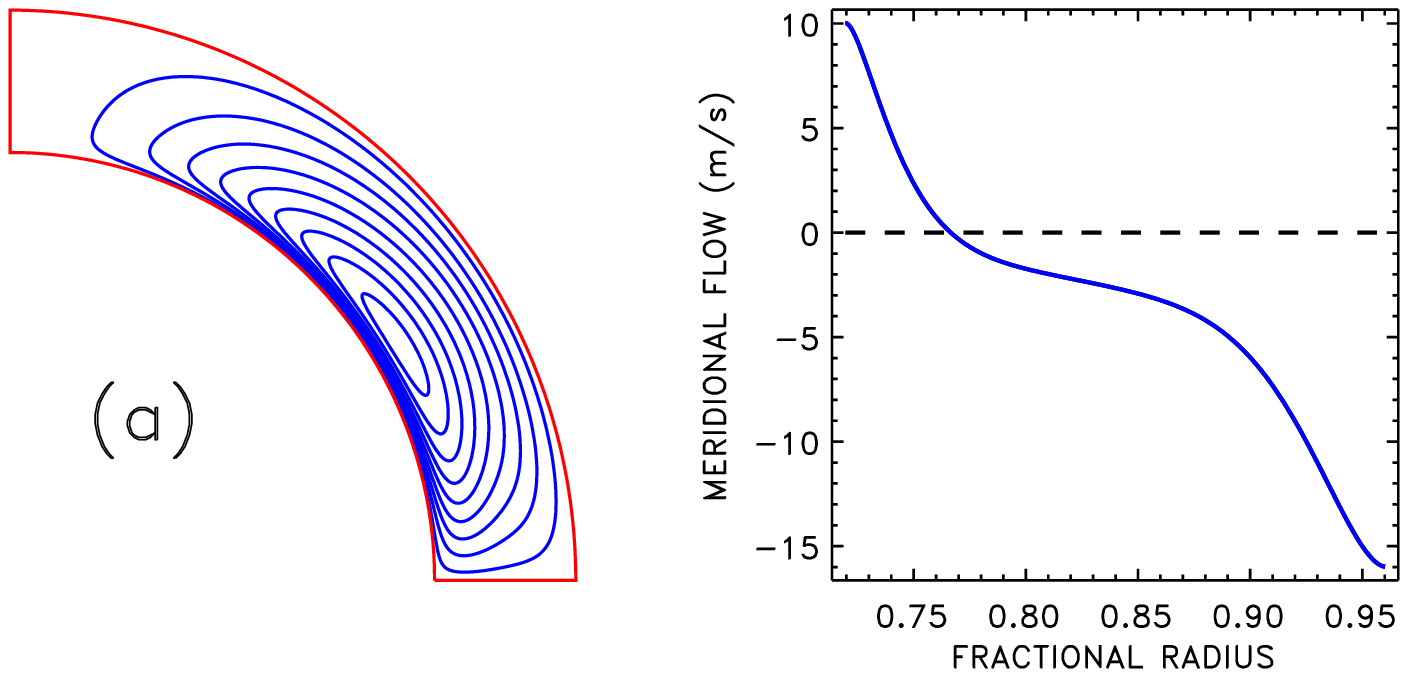}}\\[0.2truecm]
  \resizebox{\hsize}{!}{\includegraphics{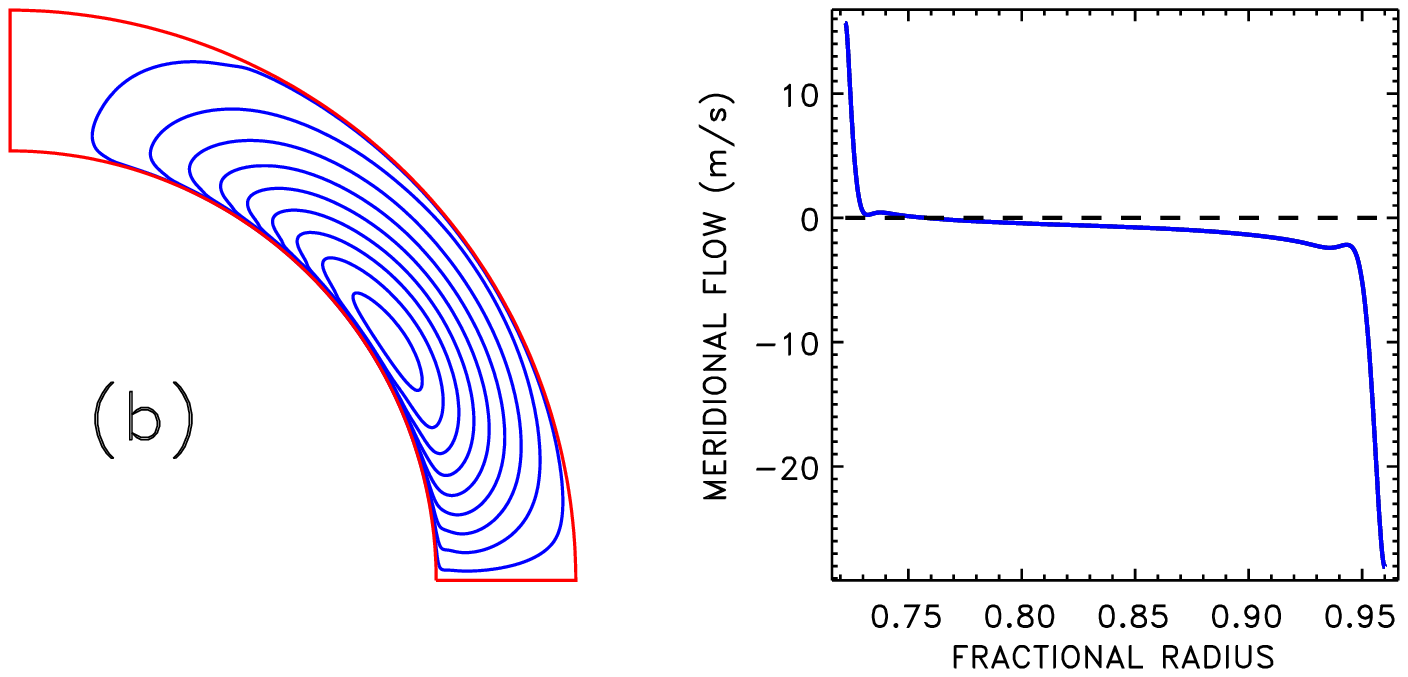}}
  \caption{Meridional flow stream lines and the flow velocity as
        function of radius at the latitude of 45$^\circ$ for
        $1M_\odot$ stars rotating with periods of 30 days (a) and
        3 days (b). Negative velocity means poleward flow.}
  \label{MFS}
\end{figure}

Violation of the Taylor-Proudman balance in the boundary layers
results in relatively large values of the sources of meridional flow
written on the right side of (\ref{12}). Accordingly, the flow
attains its largest velocities close to the boundaries.
Figure~\ref{MFS} shows the meridional flow structure for the same
computations as Fig.~\ref{f3}. The boundary layers in slow rotators
are relatively thick (KO11). The flow is smoothly distributed over
the entire thickness of the convection zone in this case. In a rapid
rotator, however, the flow is confined in thin layers near the
boundaries. This agrees with 3D simulations of
\citet{Bea08,Bea10,Bea11} who found a decrease of the meridional
flow energy with rotation rate.

\begin{figure}
  \resizebox{\hsize}{!}{\includegraphics{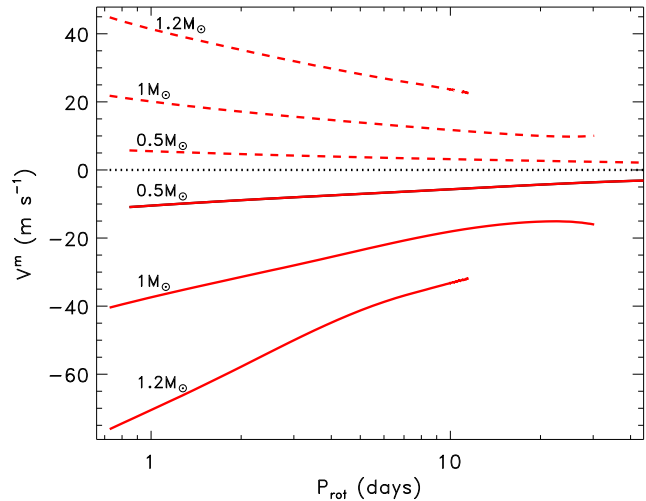}}
  \caption{Dependence of meridional velocities on the top
        (full lines) and bottom of convection zone (dashed)
        on rotation period for stars of different mass.
        The velocities are taken at the latitude of 45$^\circ$.
        Negative values mean poleward flow.}
  \label{MFD}
\end{figure}

The near-boundary flows are significant. The surface flow is
important because it is potentially observable. The flow near the
bottom of the convection zone is increasingly recognized as
important for dynamos \citep{C11}. Figure~\ref{MFD} shows how the
near-boundary flows vary with stellar mass and rotation rate in our
computations. The bottom flow is smaller but not much smaller
compared to the surface. Note, however, that the flow does not
penetrate deep beneath the convection zone. The meridional velocity
decreases rapidly with depth below the convection zone \citep{GM04}.
Similar to differential rotation, the flow amplitude increases with
stellar mass and it increases steadily with rotation rate. The flow
in the bulk of the convection zone, however, has the opposite
tendency to decrease with rotation rate (Fig.~\ref{MFS}).

\begin{figure}
  \resizebox{\hsize}{!}{\includegraphics{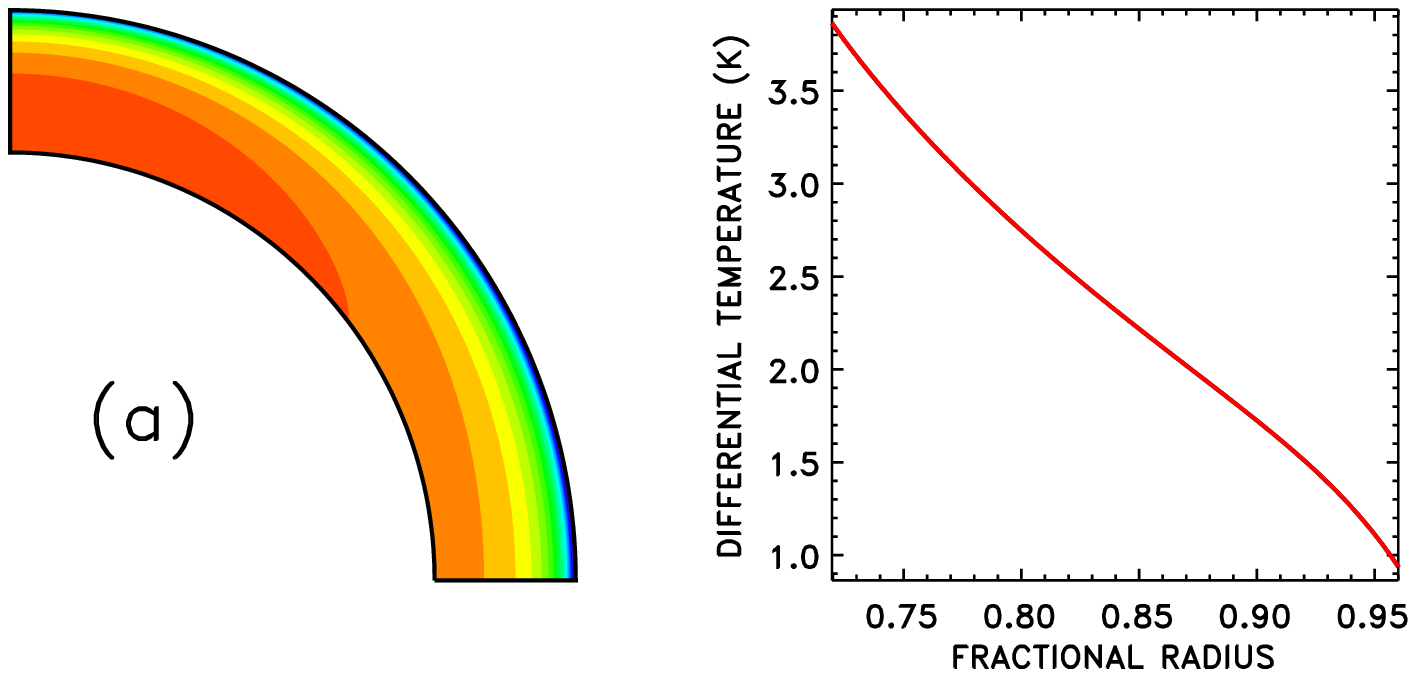}}\\[0.2truecm]
  \resizebox{\hsize}{!}{\includegraphics{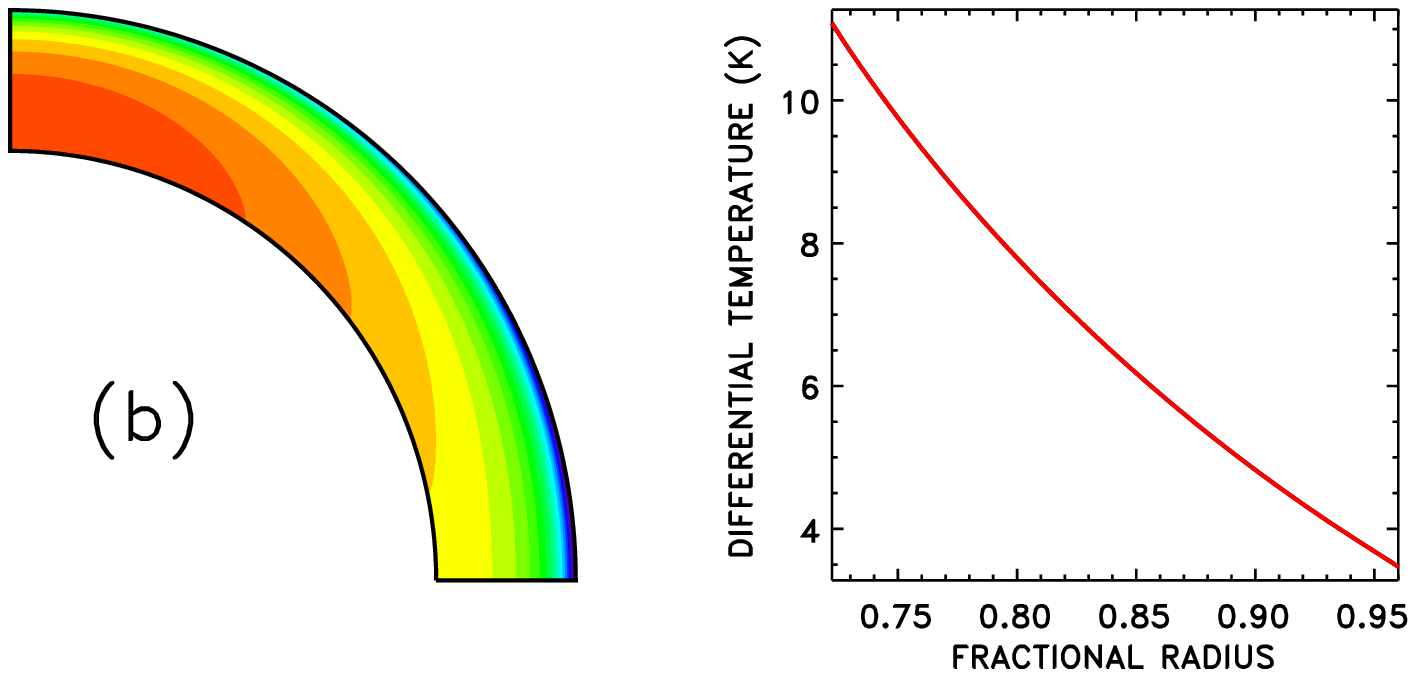}}
  \caption{Entropy isolines and depth profiles of the pole-equator
        temperature difference (\ref{deltaT}) for
        $1M_\odot$ stars rotating with periods of 30 days (a) and
        3 days (b).}
  \label{DT}
\end{figure}

The meridional flow is produced, in particular, by the effect of
thermal wind due to the temperature dependence on latitude.
Figure~\ref{DT} shows the depth profiles of the \lq differential
temperature'
\begin{equation}
    \delta T (r) = \frac{T}{c_\mathrm{p}}\left( S(r,\theta )\mid_{\theta = 0} -
    S(r,\theta)\mid_{\theta = \frac{\pi}{2}}\right)
    \label{deltaT}
\end{equation}
for the same computations as Figs.~\ref{f3} and \ref{MFS}. The
differential temperature in mean-field models is produced partly by
the dependence of thermal eddy conductivity on latitude but mainly
by anisotropy of the thermal conductivity tensor (\ref{14})
\citep{Rea05}. The anisotropy is induced by rotation and it
increases with rotation rate. Accordingly, the differential
temperature in faster rotating star of Fig.~\ref{DT} is larger.
There were many attempts to observe the small temperature difference
between the equator and poles of the sun. Recent observations of
\citet{ROM08} suggest that the solar poles are warmer than the
equator by about 2.5\,K.

\begin{figure}
  \resizebox{\hsize}{!}{\includegraphics{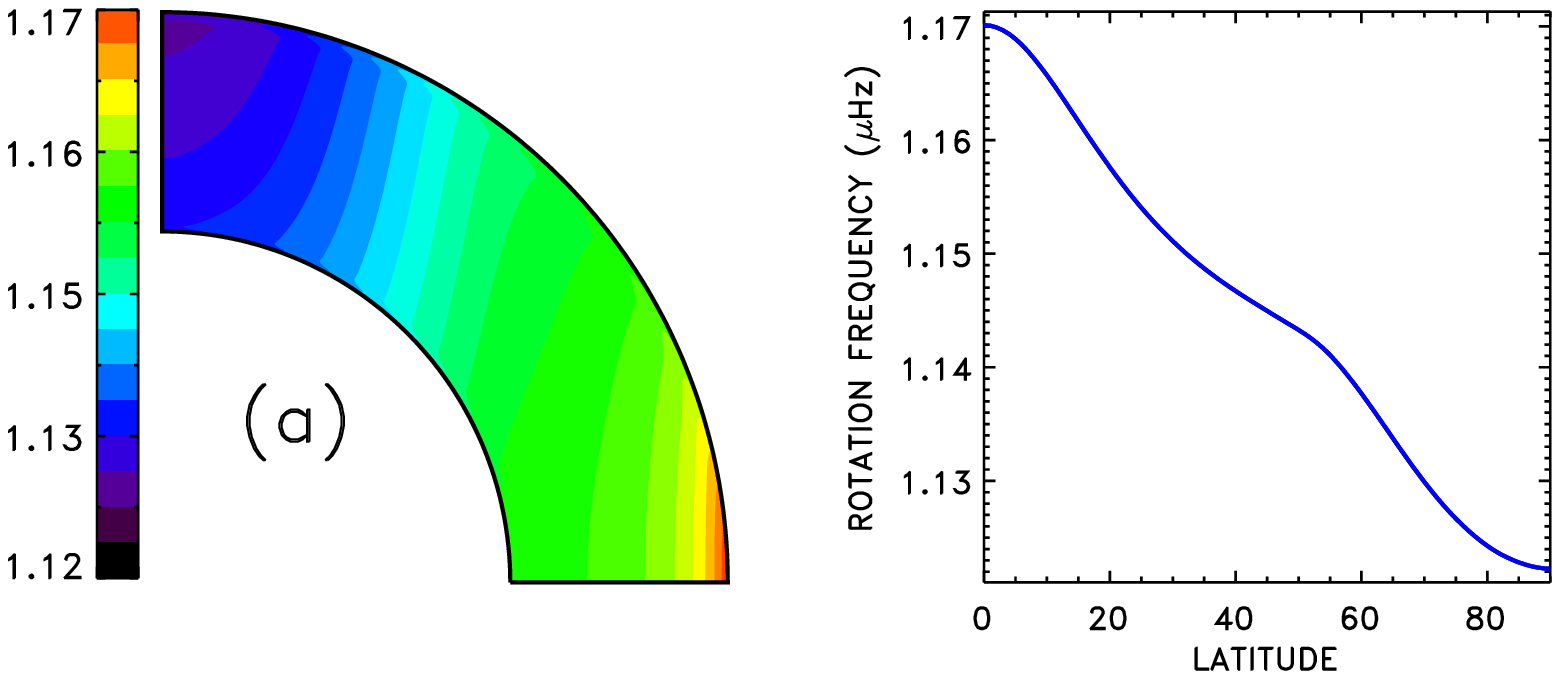}}\\[0.2truecm]
  \resizebox{\hsize}{!}{\includegraphics{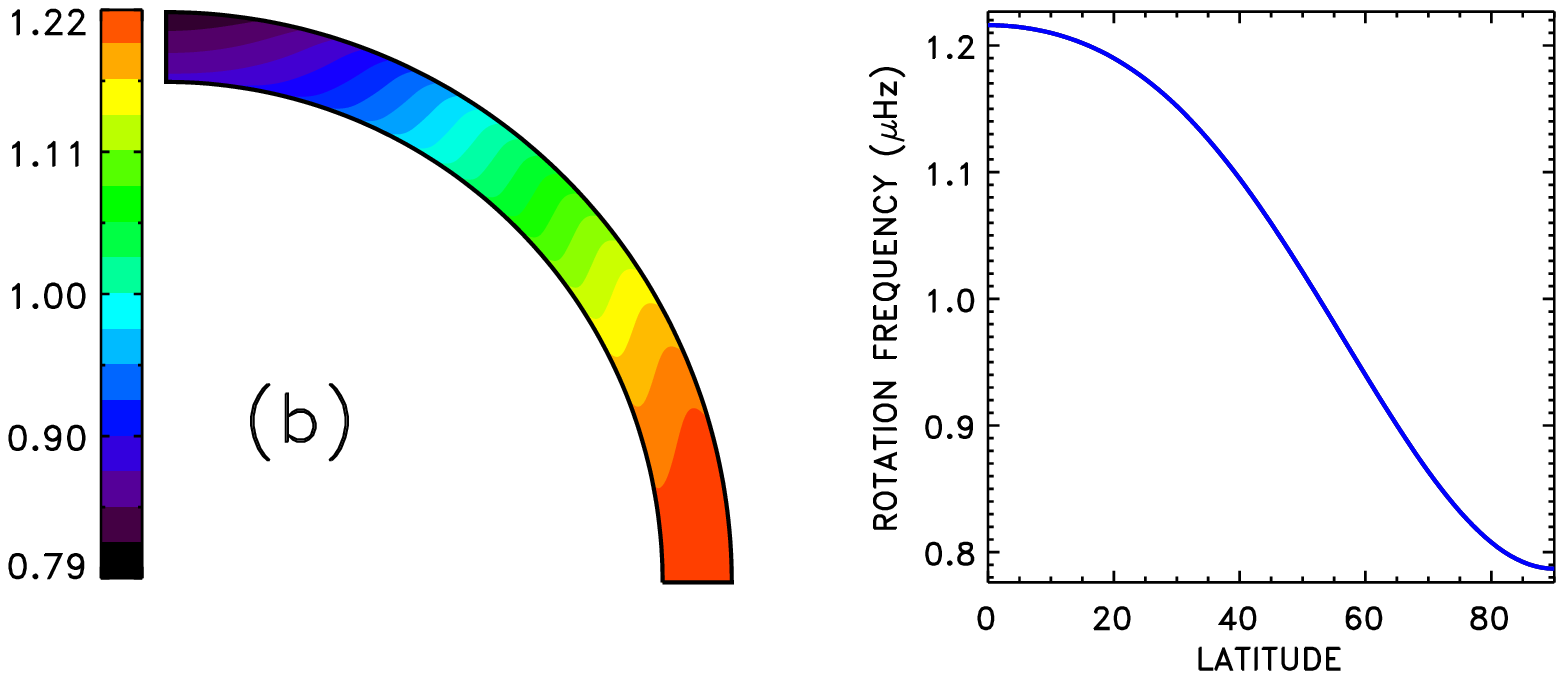}}
  \caption{Differential rotation of $0.5M_\odot$ (a) and $1.2M_\odot$ (b)
    stars, both with $P_\mathrm{rot} = 10$~days.}
  \label{f4}
\end{figure}

The anisotropy of thermal eddy conductivity is very important for
differential rotation formation. It was not possible to reproduce
the differential rotation of the sun neglecting this anisotropy
\citep{Rea05}. Reducing the anisotropy, e.g., twice results in a
change in the sense of differential rotation to an anti-solar state
(the equator rotating slower than the poles) in slowly rotating
stars. Only in this way were we able to reproduce a similar change
in the sense of differential rotation found in the 3D simulations of
\citet{Mea11} and \citet{BB11}.

Whether the differential rotation of a star belongs to the slow or
rapid rotation regime is not solely controlled by the rotation
period. Figure \ref{f4} compares the rotational states of
$0.5M_\odot$ and $1.2M_\odot$ stars with the same $P_\mathrm{rot} =
10$~days. The more massive star with $\Omega^* = 3.7$ is a typical
slow rotator while the smaller star with $\Omega^* = 49$ is not.
\section{Summary}\label{S}
Our main purpose was to predict the surface differential rotation in
terms of measurable stellar parameters. This purpose is achieved by
formulating the analytical approximation (\ref{5}) for the
differential rotation of main-sequence dwarfs as a function of their
surface temperature and rotation period (Table~1 gives the
coefficients of the approximation).

The prediction is based on the mean-field model of differential
rotation (KO11), which relies on the quasi-linear theory of
turbulent transport coefficients for rotating fluids \citep{RH04}.
The results of extensive computations for stars of different mass
and ages were processed using Gyrochronology \citep{B03,B07} and
temperature scaling for differential rotation. The scaling means
that the dependence of differential rotation on chemical composition
and mass can be combined into a common dependence on temperature.
The temperature scaling and Gyrochronology help to reduce the number
of stellar parameters on which the differential rotation depends to
just two - rotation period and temperature.

Our computations suggest that the main trend in the dependence of
differential rotation on stellar parameters is its increase with
effective temperature. The increase steepens for F-stars compared to
cooler stars (Fig.~\ref{Z}) so that the maximum surface differential
rotation, $\Delta\Omega \simeq 0.6$\,rad\,day$^{-1}$, is achieved
for the hottest F-stars we considered. There is also a weaker
dependence on rotation rate, which is not monotonous. The surface
differential rotation of young stars increases initially by about
30\% as the star ages and rotation period increases but then it
changes to a decrease with $P_\mathrm{rot}$. The maximum
differential rotation is achieved at the values of the Coriolis
number (\ref{CN}) between 10 and 20. This range of the Coriolis
number separates the regimes of fast and slow rotation. Angular
velocity distribution in slow rotators is smooth. Rapid rotators
have thin boundary layers near the top and bottom of their
convection zones where angular velocity and meridional flow vary
sharply.

One-cell meridional flow with poleward flow on the surface and a
return flow at the bottom of the convection zone was found in all
our computations. The flow attains its maximum velocity on the top
boundary. The amplitude of the flow increases with the effective
temperature, but not as much as differential rotation does, and
remains of an order of several tens of meters per second. The flow
amplitude increases smoothly with rotation rate. The structure of
the flow, however, changes considerably between cases of slow and
fast rotation. The flow is distributed smoothly over depth in the
convection zone in slow rotators. In rapid rotators, however, the
flow is concentrated in the boundary layers near top and bottom with
very weak meridional circulation in the bulk of the convection zone.
\section*{Acknowledgments}
This work was supported by the Russian Foundation for Basic Research
(projects 10-02-00148, 10-02-00391).
\bibliographystyle{mn2e}
\bibliography{kitbib}
\end{document}